\documentclass[preprint,12pt,3p]{elsarticle}
\usepackage{amssymb}
\usepackage{amsmath}
\usepackage{algorithmic}
\usepackage{graphicx}
\usepackage{textcomp}
\usepackage{xcolor}
\usepackage{caption}
\usepackage{epstopdf}
\usepackage{bm}
\usepackage{subfigure}
\usepackage{multirow}
\usepackage{mathtools}
\usepackage{float}
\usepackage{booktabs}
\usepackage[numbers]{natbib}

\usepackage{float}
\usepackage{pifont}
\newcommand{\xmark}{\ding{55}}%
\usepackage[algoruled]{algorithm2e}
\usepackage{tablefootnote}

\def\BibTeX{{\rm B\kern-.05em{\sc i\kern-.025em b}\kern-.08em
    T\kern-.1667em\lower.7ex\hbox{E}\kern-.125emX}}
\journal{}

\begin{document}
\begin{frontmatter}

\title{Non-invasive Waveform Analysis for Emergency Triage via Simulated Hemorrhage: An Experimental Study using Novel Dynamic Lower Body Negative Pressure Model \tnoteref{}}
\cortext[cor1]{Corresponding author}
\cortext[cor2]{Jointly share the second authorship}
\author[label1]{Naimahmed Nesaragi\corref{cor1}}
\address[label1]{The Intervention Centre, Oslo University Hospital,Rikshospitalet,Oslo, Norway}
\ead{naimahmed.nesaragi@gmail.com}

\author[label2,label3]{Lars Øivind Høiseth\corref{cor2}}
\address[label2]{Division of Emergencies and Critical Care, Department of Anesthesiology, Oslo University Hospital}
\ead{lars.oivind.hoiseth@hotmail.com}

\author[label1]{Hemin Ali Qadir\corref{cor2}}
\ead{hqadir2011@my.fit.edu}

\author[label2,label3,label4]{Leiv Arne Rosseland}
\address[label3]{Institute of Clinical Medicine, Faculty of Medicine, University of Oslo,Oslo,Norway}
\address[label4]{Department of Research and Development, Division of Emergencies and Critical Care, Oslo University Hospital}
\ead{lrossela@ous-hf.no}

\author[label1,label3]{Per Steinar Halvorsen}
\ead{sthalvor@ous-hf.no}

\author[label1,label5]{Ilangko Balasingham}
\address[label5]{Department
of Electronic Systems, Norwegian University of Science and Technology, Trondheim,Norway}
\ead{i.s.balasingham@ous-research.no}

\begin{abstract}
The extent to which advanced waveform analysis of non-invasive physiological signals can diagnose levels of hypovolemia remains insufficiently explored. The present study explores the discriminative ability of a deep learning (DL) framework to classify levels of ongoing hypovolemia, simulated via novel dynamic lower body negative pressure (LBNP) model among healthy volunteers. We used a dynamic LBNP protocol as opposed to the traditional model, where LBNP is applied in a predictable step-wise, progressively descending manner. This dynamic LBNP version assists in circumventing the problem posed in terms of time dependency, as in real-life pre-hospital settings intravascular blood volume may fluctuate due to volume resuscitation. A supervised DL-based framework for ternary classification was realized by segmenting the underlying noninvasive signal and labeling segments with corresponding LBNP target levels. The proposed DL model with two inputs was trained with respective time-frequency representations extracted on waveform segments to classify each of them into blood volume loss: Class 1 (mild); Class 2 (moderate); or Class 3 (severe). At the outset, the latent space derived at the end of the DL model via late fusion among both inputs assists in enhanced classification performance. When evaluated in a 3-fold cross-validation setup with stratified subjects, the experimental findings demonstrated PPG to be a potential surrogate for variations in blood volume with average classification performance, AUROC: 0.8861, AUPRC: 0.8141, $F1$-score:72.16\%, Sensitivity:79.06 \%, and Specificity:89.21 \%. Our proposed DL algorithm on PPG signal demonstrates the possibility to capture the complex interplay in physiological responses related to both bleeding and fluid resuscitation using this challenging LBNP setup.

\end{abstract}
\end{frontmatter}

\section{Introduction}
Hemorrhage with blood volume loss is one of the leading potentially preventable causes of death in trauma patients \citep{convertino2022advanced}. Hypotension is a late sign during blood volume loss due to associated physiological compensatory mechanisms. For this reason, early diagnosis of ongoing mild to moderate hemorrhage is difficult, especially in young and healthy subjects. Even invasive arterial blood pressure (ABP), exhibits poor sensitivity due to human compensatory responses \cite{chen2020estimating}. The other vital signs, including heart rate and blood oxygen saturation, also have low specificity and sensitivity for estimating blood volume loss.

Researchers have resorted to exploring various models that can artificially simulate hemorrhage. One such model is LBNP \cite{cooke2004lower,convertino2001lower,convertino2022ai,convertino2011use,van2018support}. In this model of hypovolemia, healthy volunteers are placed in an air-tight chamber to which different levels of negative pressure is applied. This retains blood in the veins of the lower extremities and pelvis, creating graded central hypovolemia. Different LBNP-levels correspond to different levels of hypovolemia. Most studies to date \cite{cooke2004lower,convertino2011use,van2018support,ji2013heart}, have applied LBNP in a predictable stepwise, progressive descending manner based on the hypothesis that \textit{"as the time elapses there is a substantial steady and linear loss of blood among the test subjects”}. When testing algorithms for classifying levels of LBNP and degree of hypovolemia, this predictability based on time of LBNP can pose a problem. For instance, in a real-life pre-hospital emergency setting,  volume resuscitation may be administrated during ongoing bleeding. We therefore propose an experimental setup with added degree of randomness in LBNP levels to avoid complete predictability by time. For the same reason, we also introduce  unequal duration at each LBNP level. Hence, our proposed experimental setup is an attempt to emulate the patient with bleeding and fluid resuscitation as may be the case in pre-hospital treatment. This experimental model is more robust in the classification of the entire dynamic LBNP trajectory for the simulated hemorrhage. 
To our knowledge, no such reliable artificial intelligence method currently exists to predict the different likelihoods among the entire trajectory of applied LBNP, and thus assist to infer the stage of hemodynamic instability independent of time.

Recently, studies on artificial intelligence (AI) based algorithms have indicated that continuous analyses of noninvasive arterial  waveform analysis (AWFA) reflect the information pertaining to the compensatory mechanisms compared to other standard vital signs \cite{convertino2022ai,chew2013haemodynamic,convertino2013estimation}. Thus,  making the earlier diagnosis of hemorrhage possible by the detection of hypovolemia prior to overt hemodynamic decompensation \cite{convertino2016compensatory}. Hence, the design of such AI-driven predictive algorithms holds the potential to reduce morbidity and mortality among patients with hemorrhage \cite{davies2020ability,summers2009validation,hatib2018machine}. In the initial screening of trauma patients, assessment is often restricted to electrocardiogram (ECG), non-invasive photoplethysmography (PPG; giving arterial oxygen saturation), and blood pressure; the first two being continuous waveforms, the latter with intermittent values. However, the PPG signal is generally considered a potential measure for variations in blood volume  because of its ability to detect intravascular volume changes \cite{convertino2022advanced,chen2020estimating,scully2012using}. Prior studies have reported that PPG-based amplitude-derived features have the potential to measure dynamic blood volume loss \cite{selvaraj2011early,shamir1999pulse,cannesson2007respiratory}. Pulse-arrival-time \cite{liang2018hypertension,mukkamala2015toward} (based on both ECG and PPG) also known as pulse transit time is used in arterial wave propagation theory for blood loss estimation \cite{djupedal2022effects}. Current early hemorrhage detection studies based on machine learning (ML) approaches rely on AWFA that mostly employs morphological changes in the features of PPG signals \cite{chen2020estimating,convertino2011use,elgendi2018toward,chen2020development,pinsky2020parsimony}. However AWFA  coupled with ML techniques and the aforementioned  techniques involves complex feature extraction to capture the subtle information for the compensatory mechanisms in the arterial waveforms. Following are the limitations involved  in the cumbersome feature extraction for PPG morphological theory and artery wave propagation theory: (i) In artery wave propagation theory, the fiducial points of each heartbeat in both ECG and PPG need to be extracted correctly \cite{elgendi2013systolic,elgendi2014detection}. (ii) This further adds  the need to have proper sync among the two modalities and also both signals have to be of high quality. (iii) It is inevitable to have optimal filtering \cite{elgendi2016optimal}. Hence, the morphological features are quite sensitive to signal quality, movement (placement) of the sensors, towards skin properties, and hence hinder the performance \cite{mejia2022comparison}.

Unlike the analysis of non-invasive signals in the time-domain, which involves beat-to-beat quantification within a sole respiratory cycle, a sequence of breaths (5-10 typically) is quantified in the spectral analysis \cite{scully2012using,pybus2019real} for estimating blood volume loss. Prior studies \cite{ji2013heart,scully2012using} that coupled LBNP experiment setup with AI have efficiently used  time-frequency (T-F) spectral methods for the assessment of blood volume loss in  awake, spontaneously breathing subjects. The present study also focuses on the assessment of two non-invasive signals viz., ECG and PPG using high- resolution transient signatures based on T-F spectral analysis to detect progressive hypovolemia in awake spontaneously breathing subjects.

The present study aimed to  (i)  determine to what extent non-invasive ECG and PPG waveforms when coupled with ML (more specifically DL) predictive analytics can classify the degree of hypovolemia in healthy volunteers undergoing LBNP with added randomness both in level and duration of each LBNP-level to reduce the effect of time, (ii) to compare the diagnostic capability of efficient T-F representation schemes with classical feature extraction methods.

\section{Materials and Methods}\label{sec5}

\subsection{  Study Population and Data Sources }
The study was approved by the regional ethics committee (REK sør-øst C/ 2019/ 649). After written informed consent, 23 healthy volunteers aged between 18 and 40 years were included in the study. Pregnancy and/or cardiovascular disease with medication were exclusion criteria.
Three-lead ECG was sampled from the Solar 8000i (GE Medical Systems) and a BioAmp/ PowerLab (ADInstruments, Bella Vista, Australia). PPG was sampled from a Masimo Radical 7 pulse oximeter, software 7.3.1.1 (Masimo Corp., Irvine, CA, USA). The sampling rate was  1000 Hz.

\subsection{Experimental Protocol }
\begin{figure}[!h]
    \centering
    \footnotesize
    \begin{tabular}{cc}
         \includegraphics[trim=4cm 1.1cm 7.5cm 3.0cm, clip=true,width=2.75in,height=2.5in]{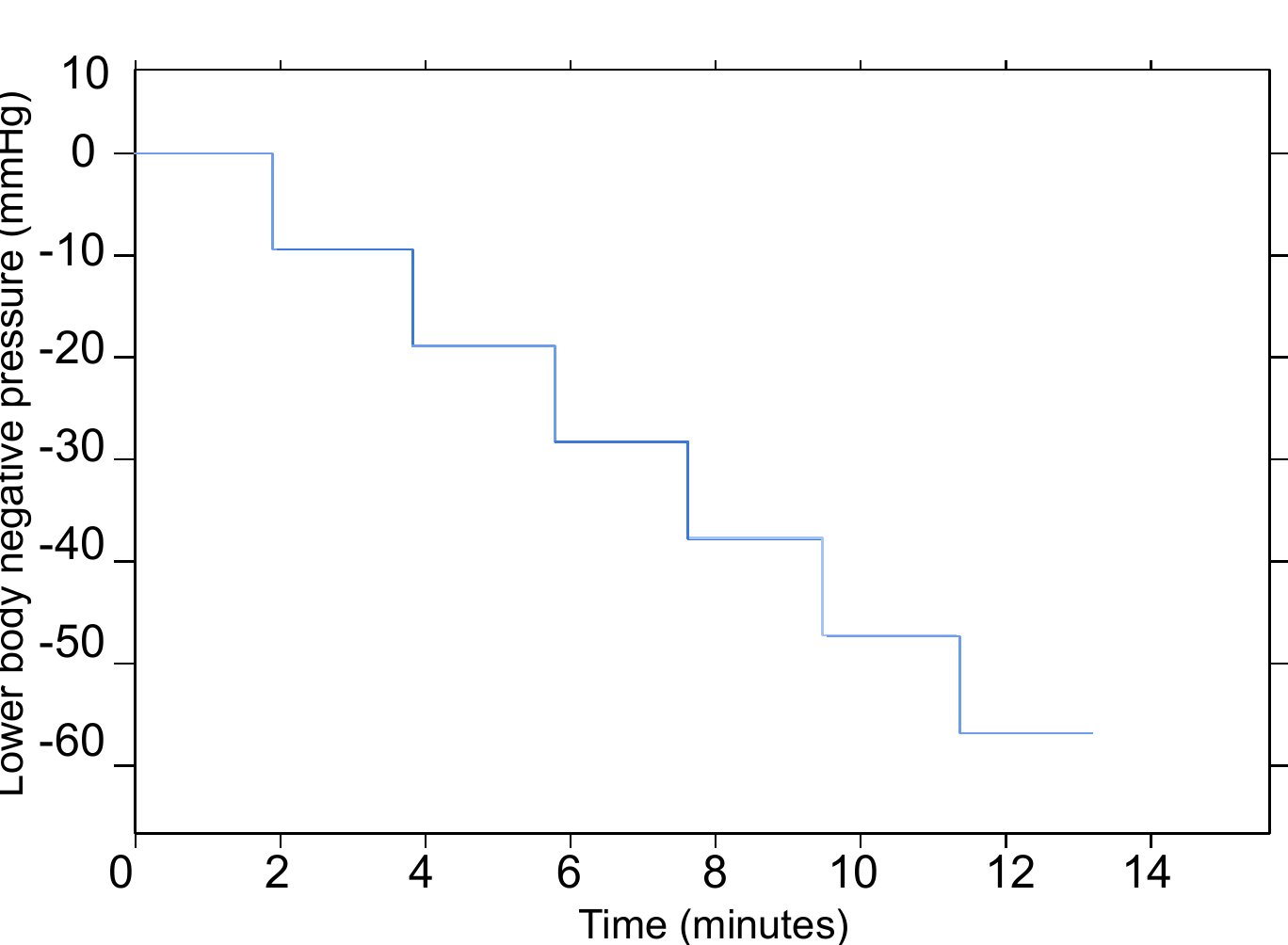} & \includegraphics[trim=2.0cm 0.9cm 21.0cm 0.0cm, clip=true, width=3.5in,height=2.5in]{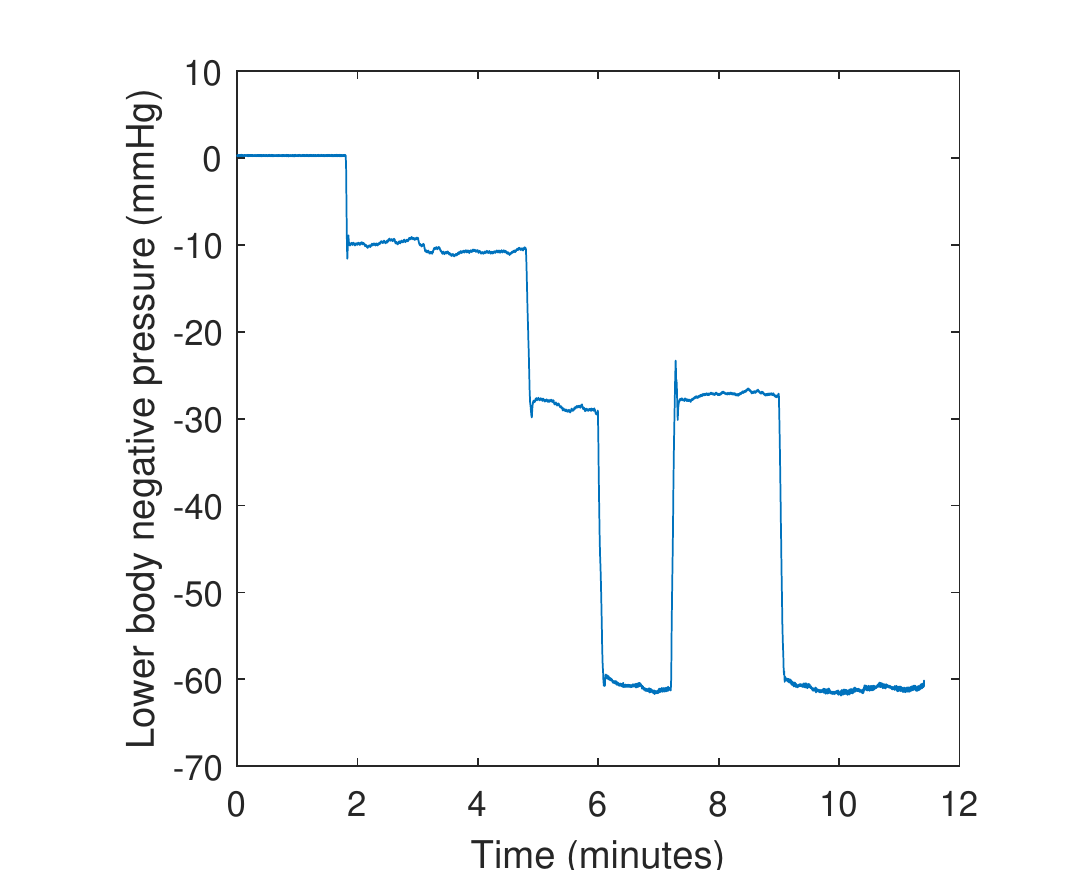} \\ 
         \footnotesize{(a)} & \footnotesize{(b)} \\
    \end{tabular}
    \caption{Illustration showing the difference between the progressive descending stepwise negative  LBNP setup (a) and random LBNP experimental protocol as used in the present study (b).}
    \label{fig2}
\end{figure}

The experimental setup used a dynamic LBNP version to study hypovolemia as opposed to the traditional model, where LBNP is applied in a predictable step-wise, progressively descending manner. Figure \ref{fig2}. shows the difference between the traditional and dynamic LBNP protocols. We refer the readers to \cite{goswami2019lower} for preliminaries and further details on traditional LBNP setup. The lower part of the volunteers’ body is subjected to a negative atmospheric pressure  applied via the pressure  chamber. Blood is drawn towards legs and pelvis, to reduce central blood volume and thus emulate hemorrhage. Before the experiment started, the subject was familiarized with the setup resting in  the supine position.  
Thereafter, the subjects were exposed to stepwise  LBNP starting at 0 mmHg with unequal and  abrupt changes in negative pressure, for  every two or three minutes. To avoid complete predictability by time, a degree of randomness was added through the experiment  as compared to the general trend of progressive descending LNBP as shown in Figure \ref{fig2}.  It is to be noted that in order to increase the total trials in this cohort study, each subject was subjected thrice to different dynamic LBNP experimental protocol. So, the resulting study  cohort had 69 LBNP trials from 23 subjects.
In the case of each subject, the experiment ended at the point of hemodynamic decompensation, indicated as a sudden decrease in arterial pressure and/ or symptoms of impending circulatory collapse such as loss of color vision (gray-out), nausea or dizziness bradycardia, or sweating \cite{techentin20191d}. Once the decompensation point was reached, the application of LBNP was released immediately to ambient pressure. 


\subsection{ML Framework for Classification}
\begin{figure}[!h]
\centering
\includegraphics[trim=0.0cm 2.7cm 0.0cm 0.0cm, clip=true, scale=0.65]{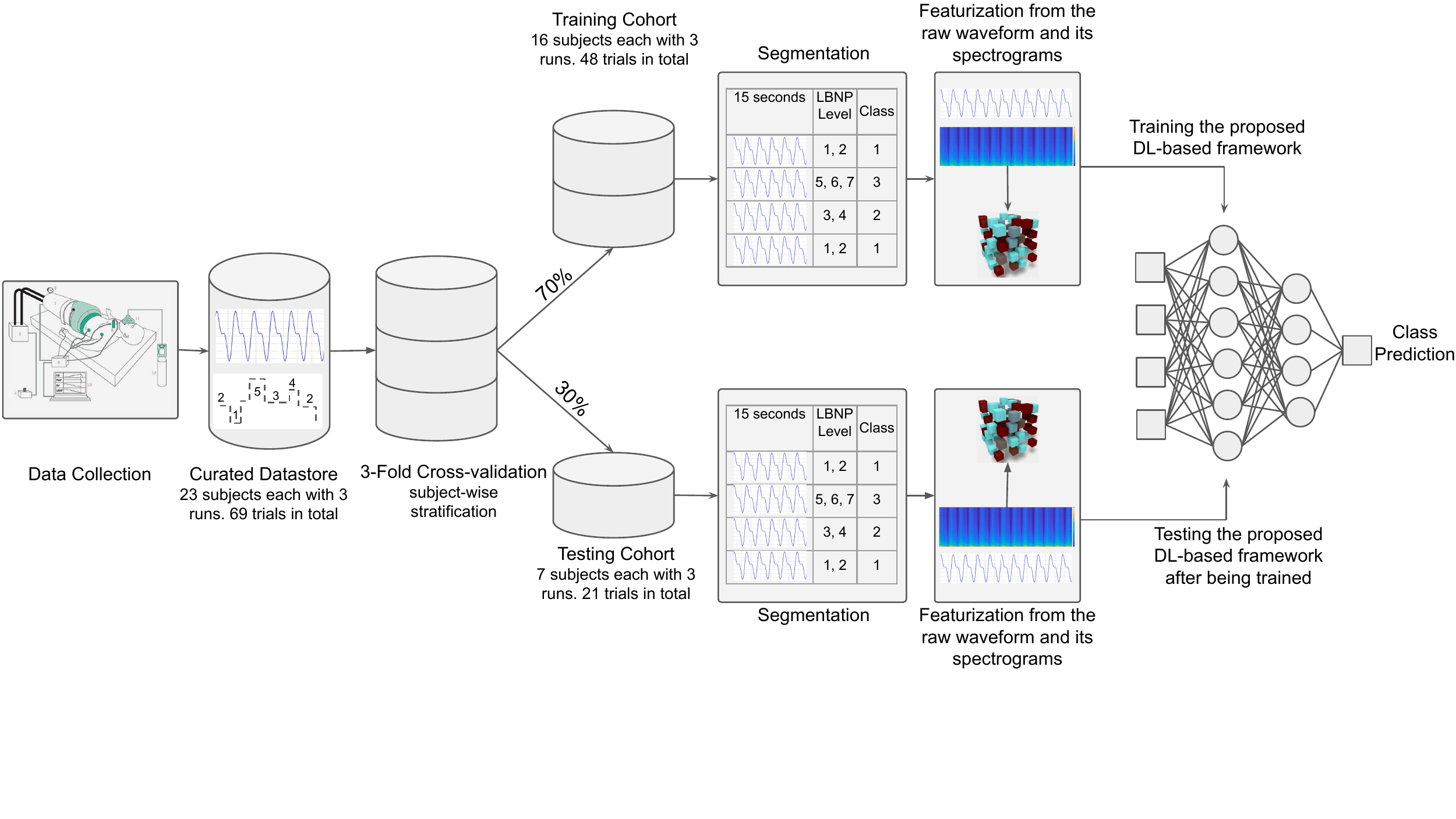}
\caption{The proposed ML-based framework consists of the following phases: \textit{Data curation} including signal preprocessing, defining training and test cohorts, and segmentation; \textit{Annotation} of the training waveform segments into different hemodynamic LBNP target levels and mapping it further to three classes; \textit{Featurization} of the non-invasive pulse waveforms (both, ECG and PPG); and \textit{Model development} for classification.}
\label{fig3}
\end{figure}

Figure \ref{fig3}. shows the overview of the proposed DL-based predictive model development at the higher level with the key phases involved in the algorithm. The proposed DL-based framework is applied for both the non-invasive signals; ECG and PPG. However, for illustration only PPG signal is considered thought the article.

\subsubsection{Data Curation}
\begin{figure}[h!]
    \centering
    \footnotesize
    \begin{tabular}{cc}
         \includegraphics[trim=2.9cm 0.0cm 16.0cm 0.0cm, clip=true, width=2.75in,height=2.5in]{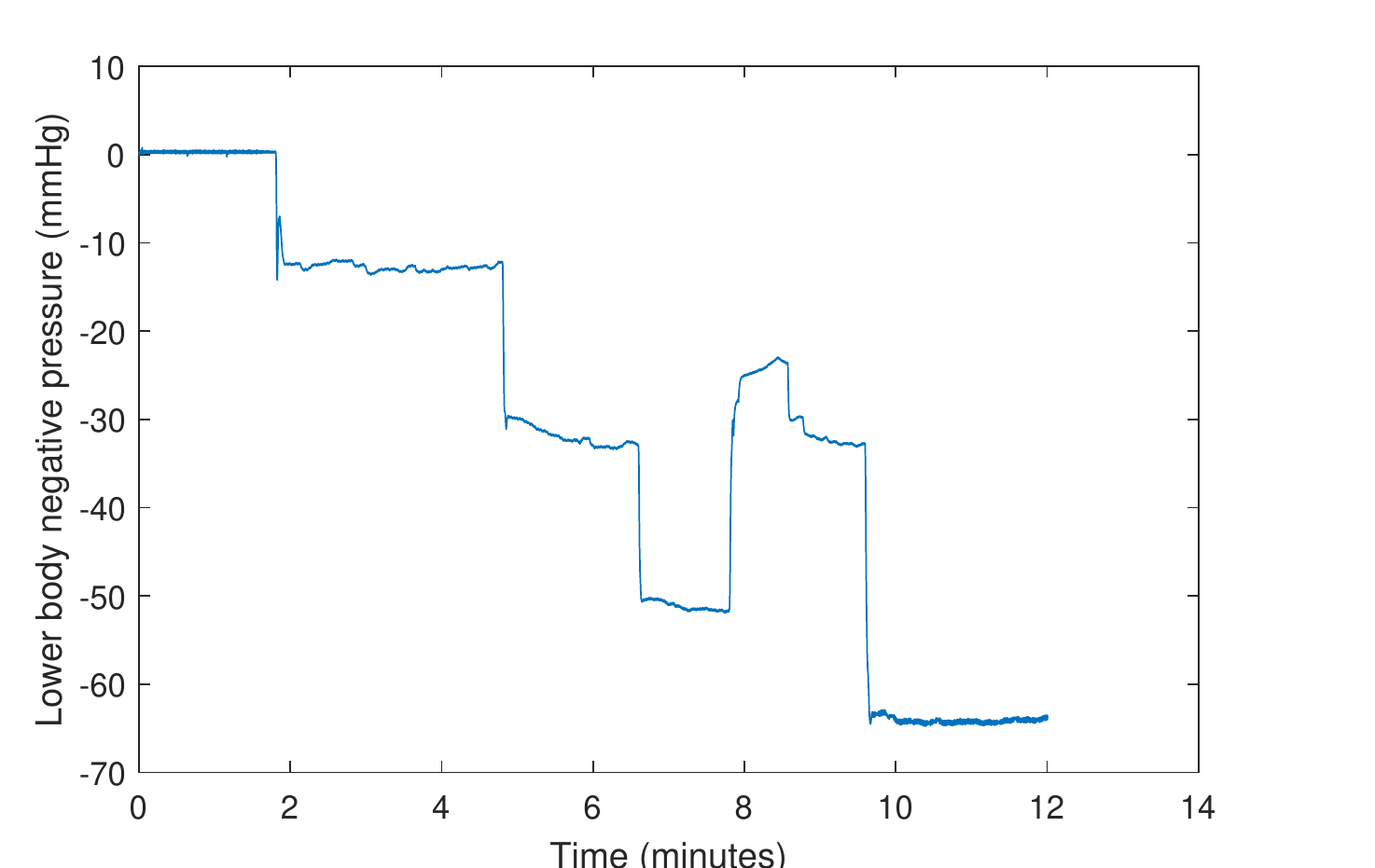} & 
         \includegraphics[trim=2.9cm 0.0cm 16.0cm 0.0cm, clip=true, width=2.75in,height=2.5in]{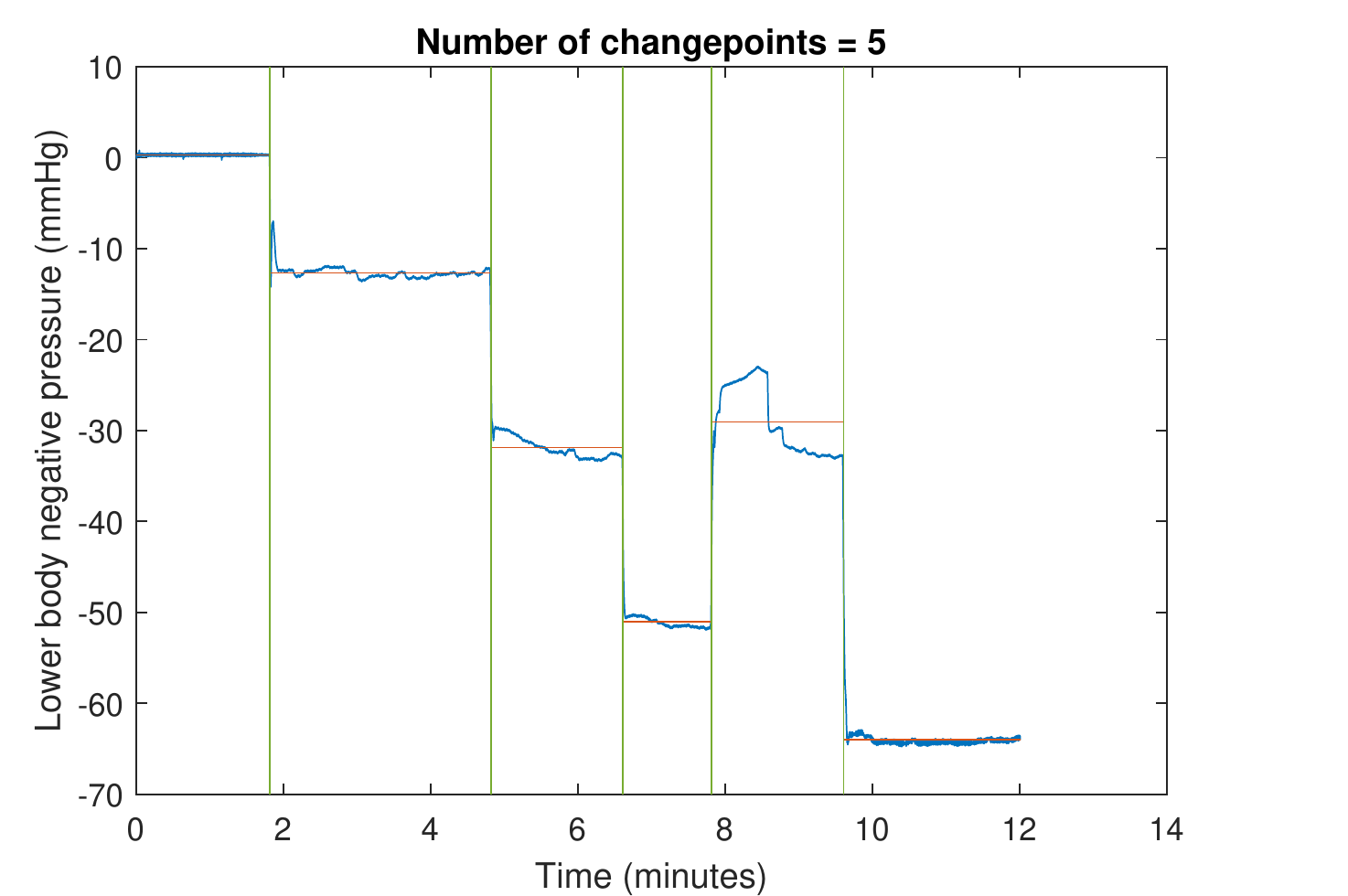}\\ 
         \footnotesize{(a)} & \footnotesize{(b)} \\
    \end{tabular}
    \caption{(a) Illustrates LBNP trial  with N=5 change points.  (b) shows the detection of endpoints marked in time  using the \textit{‘findchangepts’} function.}
    \label{fig4}
\end{figure}

The complete trajectory of the LBNP trial comprises a baseline test followed by the onset of LBNP till the pre-syncope (i.e., end-stage of LBNP) as shown in Figure \ref{fig4} (a) or completion of the protocol. The various time points defining the hemodynamic decomposition levels in the entire trajectory of the LBNP trial are marked by employing the \textit{‘findchangepts’} function in MATLAB. The \textit{‘findchangepts’} function  is based on parametric global method  as described in studies \cite{lavielle2005using,killick2012optimal} for signal changepoint detection. Hence this function can detect the abrupt changes in the LBNP trial accurately in terms of decomposition target levels as in Figure \ref{fig4} (b), thus establishing the ground truth. The mapping algorithm mentioned in Table \ref{table1} is then used to formulate the ground truth for the ternary classification among LBNP target levels. In order to predict blood volume loss via LBNP trial, it is divided into 3 classes ( Class 1 (mild):baseline to -10 mmHg; Class 2 (moderate): -20 to -30 mmHg; Class 3 (severe): over -40 mmHg). These levels correspond to estimated blood losses of 300-500 cc, 500-800 cc and greater than 800 cc respectively. This idea of association mapping between LBNP volume status to 3 classes for artificial distinction is inspired from the study performed by Soo-Yeon Ji \textit{et al.} \cite{ji2008computer}.

\begin{figure}[h!]
    \centering
    \footnotesize
    \begin{tabular}{ccc}
         \includegraphics[trim=2.9cm 0.0cm 16.0cm 0.0cm, clip=true, width=2.1in,height=2in]{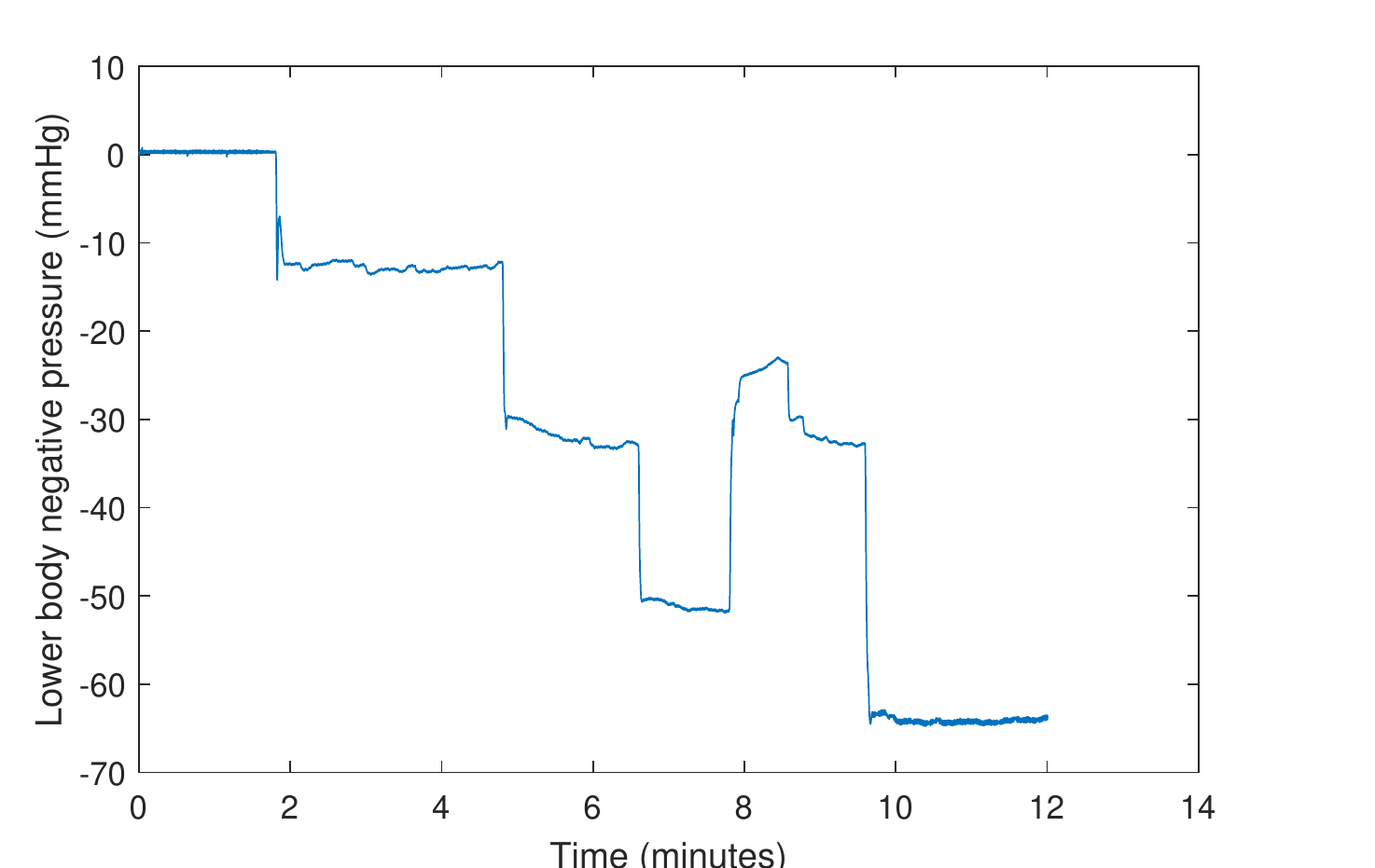} &
         \includegraphics[trim=2.9cm 0.0cm 16.0cm 0.0cm, clip=true, width=2.1in,height=2in]{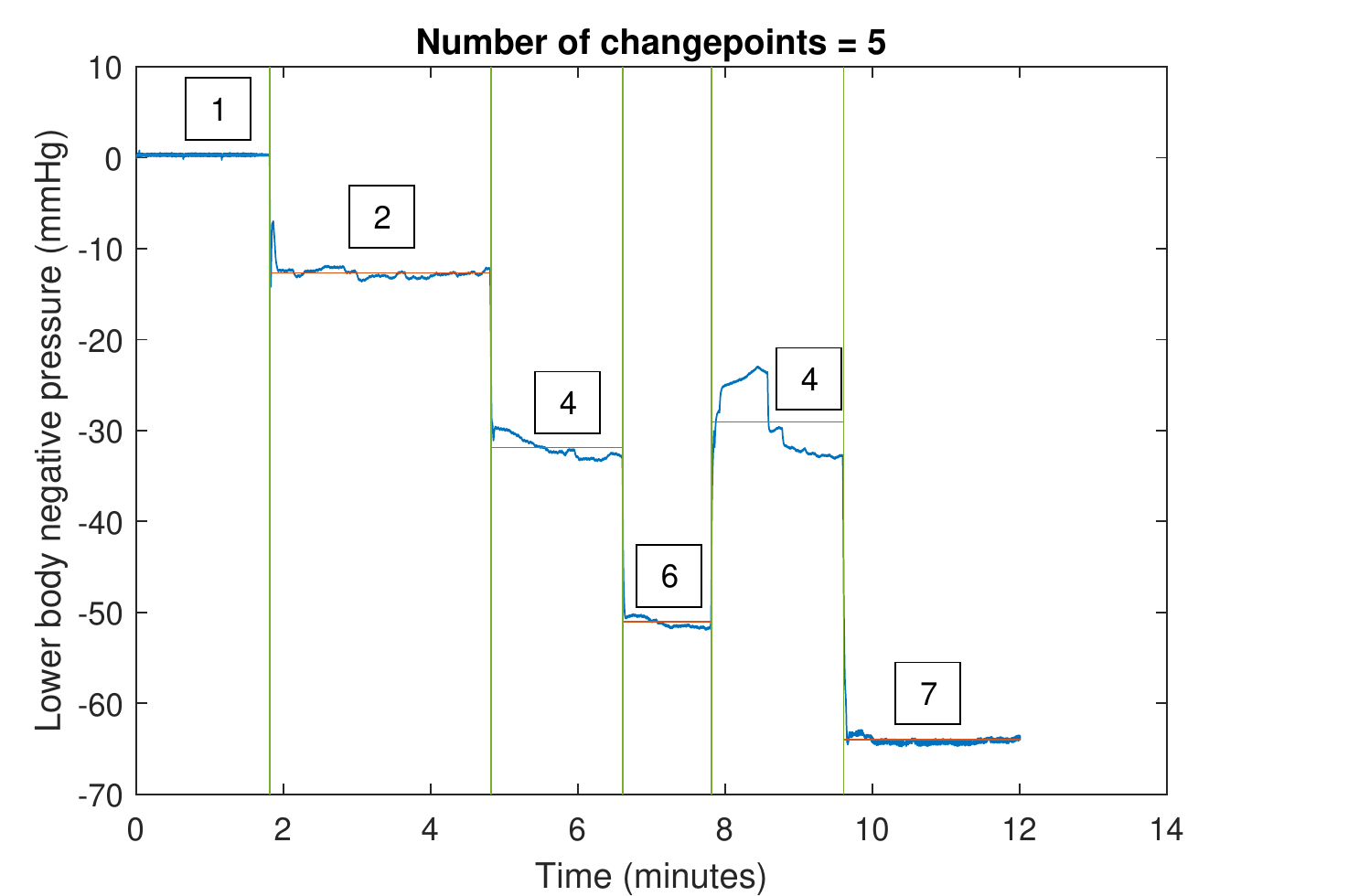} &
         \includegraphics[trim=2.7cm 0.0cm 16.0cm 0.0cm, clip=true, width=2.4in,height=2in]{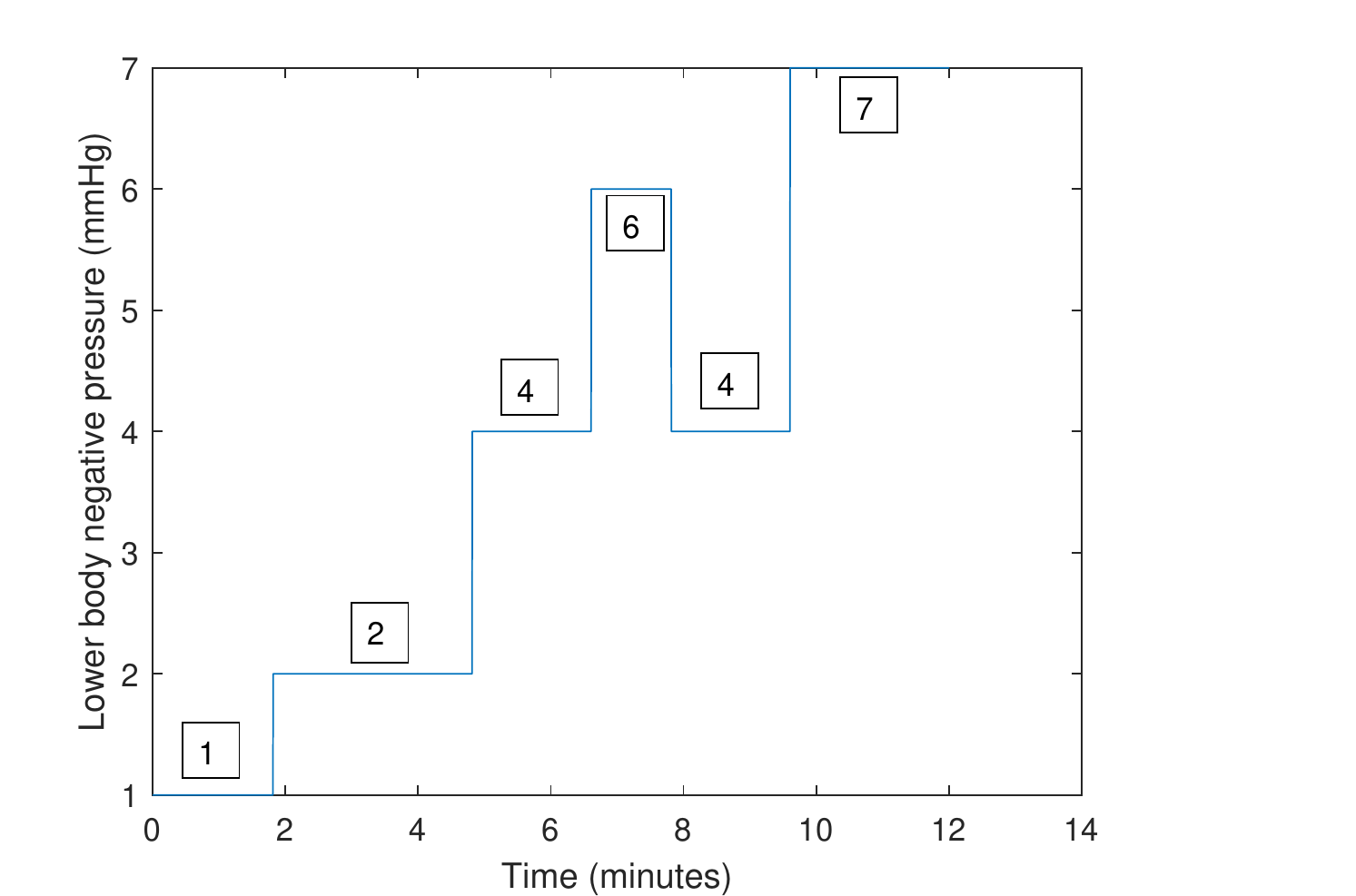}\\ 
         \footnotesize{(a)} &
         \footnotesize{(b)} &
         \footnotesize{(c)} \\
    \end{tabular}
    \caption{Definition of ground truth for the  LBNP target levels after the endpoints are marked.}
    \label{fig5}
\end{figure}

The \textit{‘findchangepts’}  function takes the number of changepoints \textit{N} as input and accurately marks the end-points in time for each LBNP target level. Once the endpoints are detected,   proper labeling  of each time point with a target LBNP level for supervised machine learning becomes feasible. The abrupt changes in the LBNP trajectory trial  are then modeled as a sequence of linear steps corresponding to the applied LBNP reference signal. As an illustration, linear and step-wise training targets for a particular LBNP trial are marked in Figure \ref{fig5} (a)-(c), together with the corresponding applied reference LBNP signal.

\begin{table}[h!]
\small
  \renewcommand{\arraystretch}{1}
  \caption{The mapping algorithm to formulate the ground truth for the LBNP target levels.}
  \label{table1} \centering
  \scalebox{1}{
\begin{tabular}{|l|l|l|}
\hline
\begin{tabular}[c]{@{}l@{}}Target\\ Level\end{tabular} & LBNP & \begin{tabular}[c]{@{}l@{}}Class\\ Definition\end{tabular} \\ \hline
\begin{tabular}[c]{@{}l@{}}1\\ 2\end{tabular}          & \begin{tabular}[c]{@{}l@{}}‘  0   mmHg'\\ ‘ -10 mmhg'\end{tabular}                         & Class 1                                                    \\ \hline
\begin{tabular}[c]{@{}l@{}}3\\ 4\end{tabular}          & \begin{tabular}[c]{@{}l@{}}‘ -20 mmHg'\\ ‘ -30 mmHg'\end{tabular}                         & Class 2                                                   \\ \hline
\begin{tabular}[c]{@{}l@{}}5\\ 6\\ 7\end{tabular}      & \begin{tabular}[c]{@{}l@{}}‘ -40 mmHg'\\ ‘ -50 mmHg'\\ ‘ \textless{} -60 mmHg'\end{tabular} & \multicolumn{1}{c|}{Class 3}                             \\ \hline
\end{tabular}}
\end{table}
Once the ground truth is established a supervised DL-based framework is formulated to predict and classify the complete trajectory run of the LBNP reference signal. This is achieved by segmenting the underlying non-invasive waveform of each subject into equal segment lengths of 15 seconds with an overlap of 10 seconds. Physiologically, a segment length of 15 seconds duration is chosen so that it captures several (12-15) heartbeats comprising at least one respiratory cycle \cite{convertino2022ai,techentin20191d}. Each such waveform segment is associated with one of the three class definitions as ground truth defined in Table \ref{table1}, and is treated as an individual sample of observation for training the proposed model. This results in a large number of observations as training samples from the original subjects with a limited number.

Next, feature extraction is performed on these waveform segments using time-frequency analysis on both non-invasive signals. The ML models are then trained with the derived feature set to classify each of 15 seconds waveform segments into either of the three classes.

\subsubsection{Classification Model}
\label{sec:class_model}
\begin{figure}[H]
\centering
\includegraphics[trim=0.0cm 1.0cm 0.0cm 0.0cm, clip=true, scale=0.65]{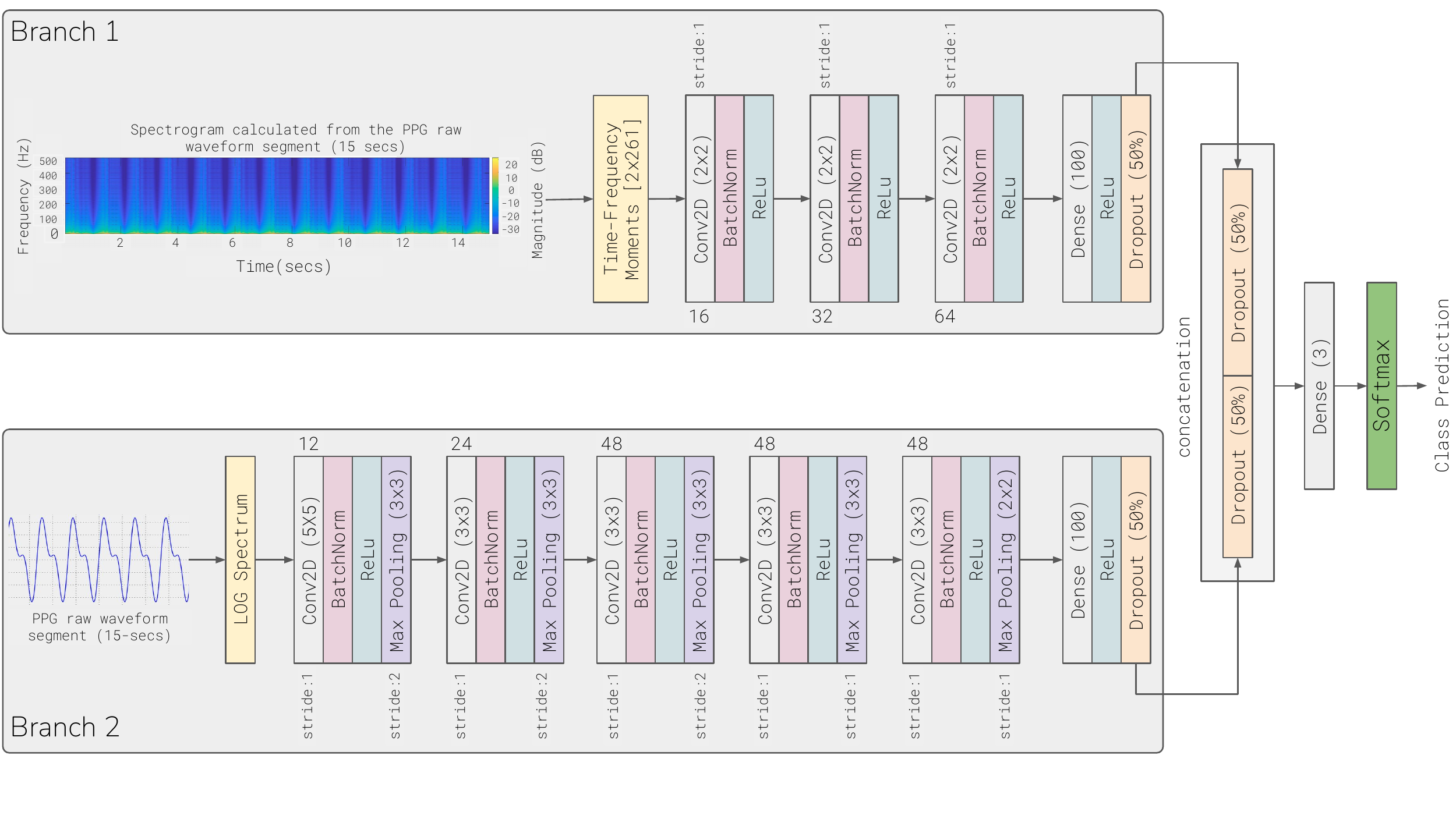}
\caption{Network architecture of the proposed unified DL-based framework}
\label{fig6}
\end{figure}

A unified DL-based model with multiple (two) inputs/branches is designed for the desired ternary classification task. The structure details with various layers in the respective branches of the proposed network architecture for the unified model is presented in Figure \ref{fig6}. The designed unified model is constructed by training a  mixture of two unique T-F representations for the respective branches, viz., T-F moments being fed to the upper input (Branch 1)  of the network architecture as shown in Figure \ref{fig6}. The detailed explanation of the feature extraction in terms of T-F moments derived from the spectrograms of given waveform segments is presented in the subsequent $section$ \textit{2.3.3}. The lower input (Branch 2) of the network is initially fed with a raw waveform segment which is further converted into a logaritham scale-based  2-D spectrogram by custom-defined \textit{'log spectrum layer'} to train the subsequent 2D-CNN layers. The need and advantage of such custom-defined \textit{'log spectrum layer'} are also described in $section$ \textit{2.3.3}.


\subsubsection{Featurization}
\begin{figure}[H]
    \centering
    \footnotesize
    \begin{tabular}{c}
         \includegraphics[trim=6.35cm 1.2cm 4.85cm 0.0cm, clip=true, scale=0.27]{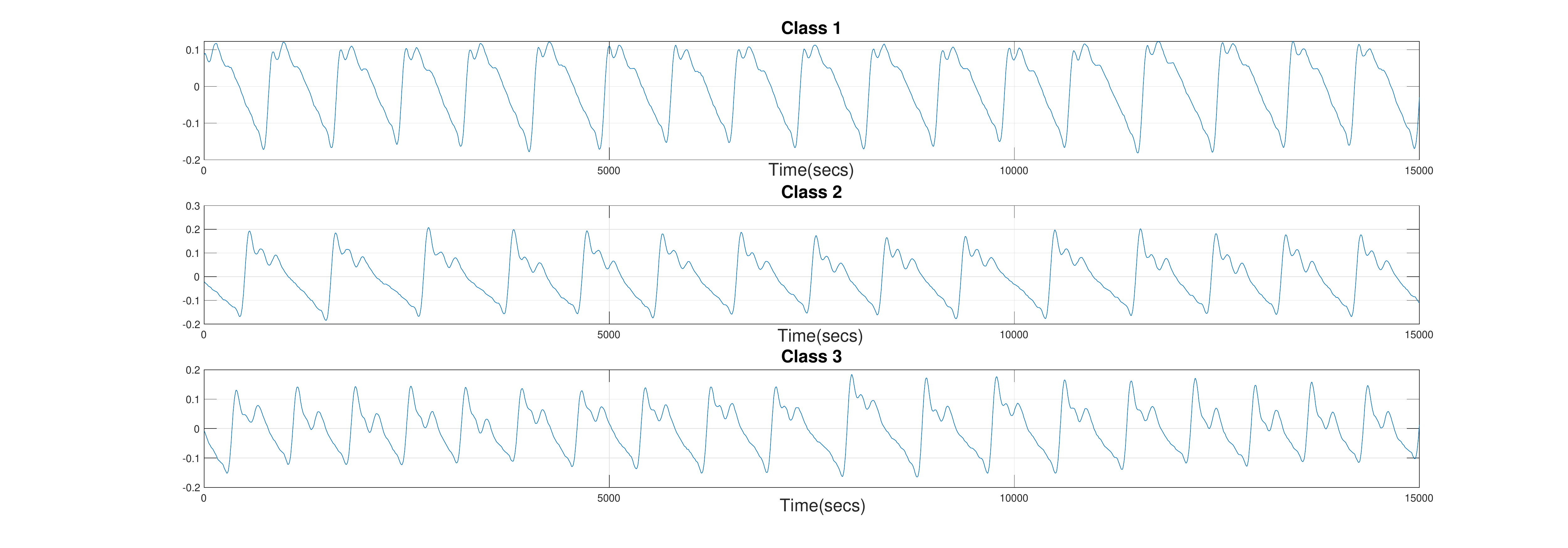} \\ 
         \footnotesize{(a) } \\
         \includegraphics[trim=5.35cm 1.3cm 4.0cm 0.0cm, clip=true, scale=0.27]{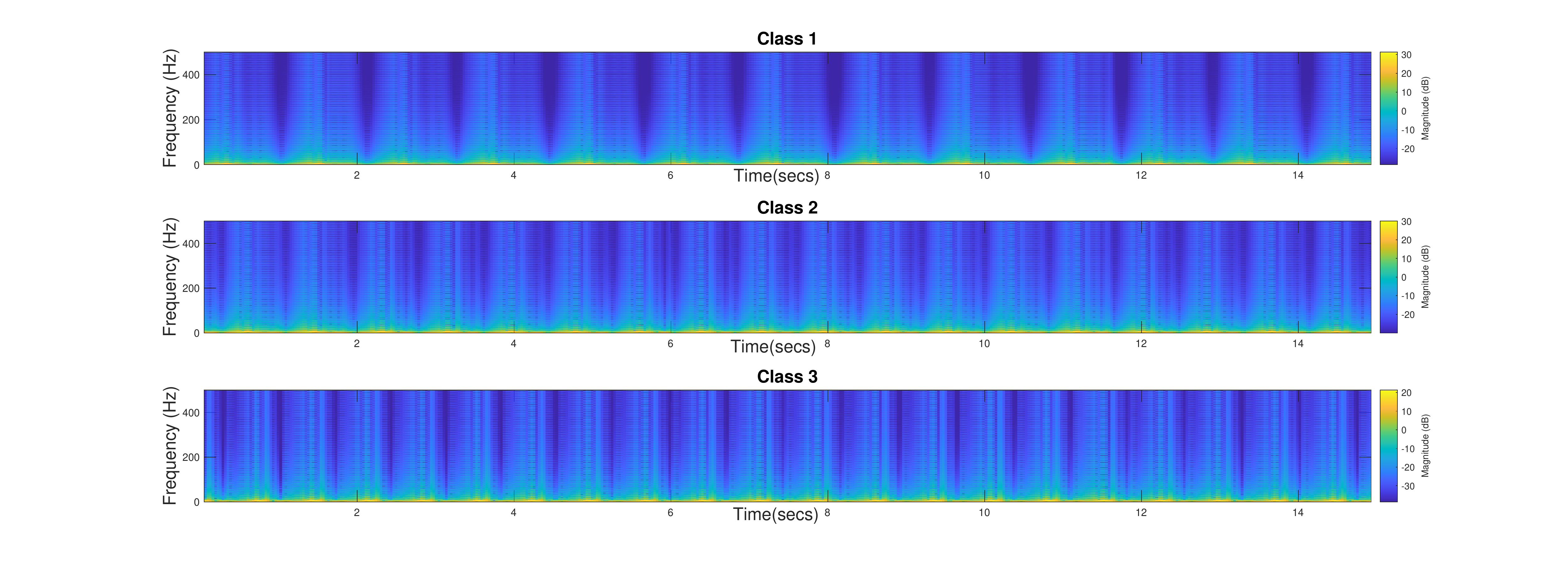} \\
         \footnotesize{(b) } \\
         \includegraphics[trim=5.35cm 1.3cm 4.0cm 0.0cm, clip=true, scale=0.27]{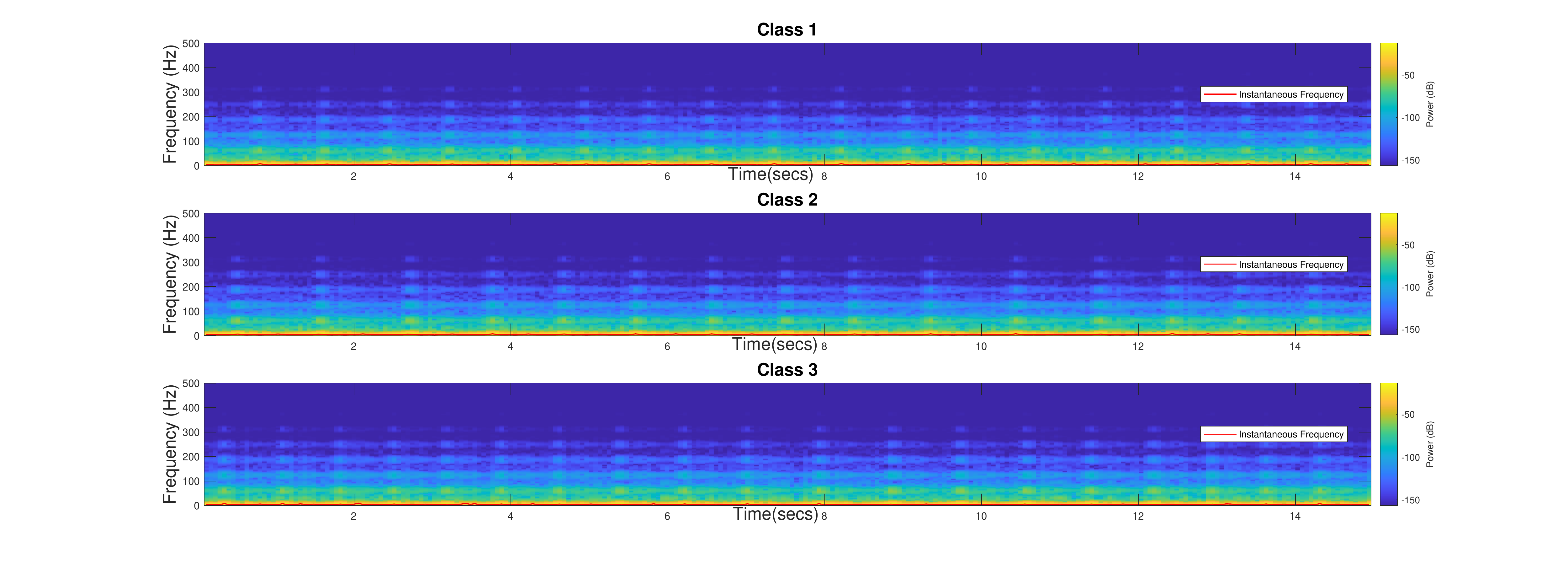} \\ 
         \footnotesize{(c) } \\
         \includegraphics[trim=5.35cm 1.3cm 4.8cm 0.0cm, clip=true, scale=0.27]{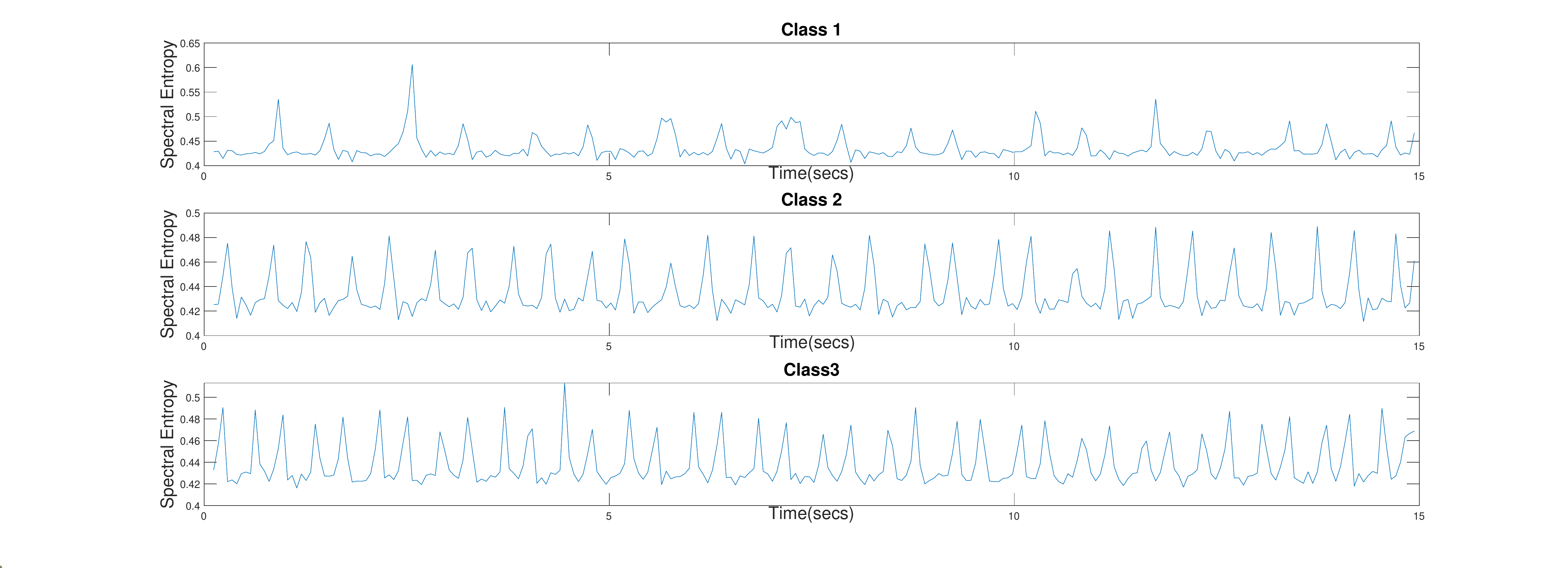} \\
         \footnotesize{(d) }
    \end{tabular}
    \caption{(a) Illustration of raw PPG waveform segments for the three different classes. (b) Spectrograms derived from the raw PPG waveform segments shown in (a). (c) Instantaneous frequencies and (d) Spectral entropies computed from the spectrograms shown in (b).}
    \label{fig:fig7}
\end{figure}

In this work, efficient feature extraction is realized by exploring T-F analysis of the underlying non-invasive signal. Thus the time-series signal in the  1-D domain as depicted  in Figure \ref{fig:fig7}(a)  is converted into the  2-D real plane (see Figure \ref{fig:fig7}(b)) to extract transient signatures. In literature, spectrogram-based T-F analysis, has been extensively employed with recurrent and convolutional neural networks (CNNs) to extract diagnostic signatures for various clinical applications \cite{8743898,elola2022beyond}. However, dimensional reduction of the resultant time-frequency feature space can reduce the complexity of the  algorithm and improve the classification performance with increased intelligibility for decision-making. 
This can be realised in practice  by extracting T-F moments from the spectrograms. The present study explores two such moments in the T-F domain viz., spectral entropy (SE) and instantaneous frequency (IF) \cite{boashash1992estimating,boashash1992estimating1}.
Figures \ref{fig:fig7}(c) and 6(d) illustrates the differences between IF and SE for windowed hemodynamic LBNP regions of the typical PPG signal. These T-F moments derived from the spectrograms of the noninvasive signal provide the best granular information of the two worlds, both fine-granularity and coarse-granularity. This can be explained as follows. For fine-granularity, first the given raw waveform segment of 15 seconds, sampled at 1000 Hz (corresponds to 15000 discrete time series) is first converted to length $N$ of radix-2. i.e. $N=2^n$, where n is a positive integer. This is achieved either by truncating the discrete time series or by padding it with zeros so that $N=2^n$. In our case, for a discrete time-series with an initial length of 15000, the next close radix-2 number is $N=2^{14}$=16384. So the time-series segment is padded  with zeros to increase its length to 16384. Then for finer-granularity the time series is binned with a window function of length 64 to form 261 time windows. Later, for coarse granularity, the central moment from power-spectrogram is computed, which corresponds to the center of the time windows.

The IF is the time-dependent frequency of a signal under interest and is computed as the first moment from the power spectrum that represents the spectral density resulting from short-time Fourier transforms as defined in equation (1), where  $P(t, f)$ is the power spectrum of the time-window \cite{boashash1992estimating,boashash1992estimating1,buttkus2012spectral}.

\begin{equation}
I F(t)=\frac{\int_{-\infty}^{\infty} f P(t, f) d f}{\int_{-\infty}^{\infty} P(t, f) d f}
\end{equation} For a given non-invasive signal (sampled at 1000 Hz)  waveform segment with a duration of 15 seconds, a feature vector of 261 lengths is obtained by computing spectrograms over 261-time windows. The output values are IF in time i.e \textit {IF(t)}, corresponds to the center of the time windows.

The SE combines the knowledge of spectrogram-based spectral density analysis with the information-theoretic measure- Shannon entropy \cite{boashash1992estimating,boashash1992estimating1,buttkus2012spectral}. SE reflects the degree of randomness (uncertainty) or the regularity (deterministic patterns) in the signal of interest. A spiky or random signal has low  SE, while deterministic signals like white noise with a flat spectrum have higher SE values. The estimation procedure of SE is similar to  IF and uses 261-time windows for the corresponding  non-invasive signal waveform segment with duration of 15 seconds. 
However, SE considers the normalized power distribution in the frequency domain as a probability distribution of the signal and calculates its Shannon entropy. Therefore, the calculated Shannon entropy is contextually  known as the SE of the signal. Given a T-F power spectrogram   \textit{P(t, f)}, the probability distribution at frequency point \textit{n, n=1,…,N;} and time \textit{t}, $0 \leq t \leq T$;  denoted as \textit{p(t, n),} is
\begin{equation}
p(t, n)=\frac{P(t, n)}{\sum_f P(t, f)},
\end{equation}where $f \in$ [0,$f_s$/2], and $f_s$ = 1000 Hz, sampling frequency. Then SE at time \textit{t}, denoted as $S(t)$, is given as \cite{boashash1992estimating,boashash1992estimating1,buttkus2012spectral}:

\begin{equation}
S(t)=-\sum_{n=1}^N p(t, n) \log _2 p(t, n).
\end{equation}

The custom definition of \textit{'log spectrum layer'} layer uses the  \textit{'dlstft'} function in MATLAB for computing short-time Fourier transforms that inherently support automatic backpropagation. The need and the advantage of such a custom realization of the network layer can be explained as follows. When any pre-processing steps, involving signal processing are performed outside the DL network, then predictions might differ due to different pre-processing settings in comparison to those used in training of the network \cite{salvi2021impact}. This can impact the  performance of the network (to be poorer than expected). Placing the pre-processing computations (in this case spectrograms) inside the network as a layer results in self-contained model and simplifies the pipeline for deployment with efficient handling of storage \cite{zhou2017optimized,karimi2019deep,salvi2020fully}.  The logarithm-based scale of the spectrogram is  considered in training the deep networks because it acts like a  
dynamic range compressor. This boosts the representation scheme having values with small magnitudes (amplitudes) but still carrying important information.

\section{Experiments and Results}\label{sec5}

\subsection{Dataset Stratification for Cross-validation}
The proposed framework performed predictive analytics on the given  non-invasive wave-form segments of both ECG and PPG signals from the patient records to determine the degree of risk in hypovolemia development  by classifying the reference LBNP trajectory into 3 classes. The experimental study was performed using  three-fold cross-validation based on a patient-wise stratification scheme, with each fold containing a unique $\sim$30\% of the entire dataset. i.e., The model is trained and developed using $\sim$70\% (16 subjects) of total data and the remaining  $\sim$30\% (7 subjects) data is considered for validation. Thus, the total data comprising 23 subjects is divided into stratified, three unique training and test sets containing 16 and 7 subjects respectively to perform three-fold cross-validation.

\begin{table}[h!]
\small
  \caption{Sample distribution of waveform segments for model training and validation in 3-fold cross-validation}
  \label{table} \centering
  \scalebox{.9}{
\begin{tabular}{|llll|}
\hline
\multicolumn{4}{|c|}{Training}                                                                      \\ \hline
\multicolumn{1}{|l|}{}        & \multicolumn{1}{l|}{Fold 1} & \multicolumn{1}{l|}{Fold  2} & \multicolumn{1}{l|}{Fold  3} \\ \hline
\multicolumn{1}{|l|}{Class 1} & \multicolumn{1}{l|}{1263}      & \multicolumn{1}{l|}{1215}        &    \multicolumn{1}{l|}{1174}    \\ \hline
\multicolumn{1}{|l|}{Class 2} & \multicolumn{1}{l|}{1843}       & \multicolumn{1}{l|}{1866}        &   \multicolumn{1}{l|}{1955}     \\ \hline
\multicolumn{1}{|l|}{Class 3} & \multicolumn{1}{l|}{1997}       & \multicolumn{1}{l|}{2079}        &    \multicolumn{1}{l|}{2055}    \\ \hline
\multicolumn{1}{|l|}{Total}   & \multicolumn{1}{l|}{5103}       & \multicolumn{1}{l|}{5160}        &    \multicolumn{1}{l|}{5184}    \\ \hline
\multicolumn{4}{|c|}{Testing}                                                                       \\ \hline
\multicolumn{1}{|l|}{}        & \multicolumn{1}{l|}{Fold 1} & \multicolumn{1}{l|}{Fold  2} & \multicolumn{1}{l|}{Fold  3} \\ \hline
\multicolumn{1}{|l|}{Class 1} & \multicolumn{1}{l|}{301}       & \multicolumn{1}{l|}{333}        &      \multicolumn{1}{l|}{365}   \\ \hline
\multicolumn{1}{|l|}{Class 2} & \multicolumn{1}{l|}{601}       & \multicolumn{1}{l|}{582}        &    \multicolumn{1}{l|}{598}     \\ \hline
\multicolumn{1}{|l|}{Class 3} & \multicolumn{1}{l|}{633}       & \multicolumn{1}{l|}{582}        &      \multicolumn{1}{l|}{546}   \\ \hline
\multicolumn{1}{|l|}{Total}   & \multicolumn{1}{l|}{1535}       & \multicolumn{1}{l|}{1497}        &      \multicolumn{1}{l|}{1509}   \\ \hline
\end{tabular}}
\end{table}

Defining stratified cohorts  in terms of individual subjects is very much  important during training and testing, since over-fitting (high variance)  is usually observed in experiments where validation waveform segments are selected from the pool of all subjects.
It is to be noted that each subject is subjected thrice to the LBNP experimental protocol. So, the resulting training cohort included training waveform segments with 48 LBNP trials from 16 unique subjects and the testing cohort included  waveform segments with 21 LBNP trials from the remaining unique 7 subjects. The sample distribution of waveform segments in the three-fold validation setup is listed in Table 4. Further it is worth mentioning that the overlap of 10 seconds duration is performed only during training and is omitted during the segmentation of test subjects, to keep the ratio of 3:1 among the waveform segments of train and test cohorts. 

\subsection{Model Design and Training}
The proposed DL-based classification model was developed using the DL toolbox in MATLAB with NVIDIA GeFore GTX 1080Ti. In a 3-fold cross-validation setup, the proposed model was trained on subject-wise stratified three-folds  and the optimized hyper-parameters that minimize the cross-validation loss are listed in Table \ref{table3}. The optimal network hyper-parameters and training options were obtained by performing Bayesian optimization using \textit{'Experiment Manager'} in MATLAB. An objective function was formulated for the underlying Bayesian optimization on model hyper-parameters that intend to maximize the $F1$-score.

\begin{table}[h!]
\small
\centering
\caption{Optimized model hyper-parameters with training option  values.}
\scalebox{.9}{
\begin{tabular}{|l|l|}
\hline
\begin{tabular}[c]{@{}c@{}c@{}}  \textbf{Model hyper-parameters}  \\ \textbf{ \& Training options}  \end{tabular}  & \textbf{Values} \\ \hline
$L2\_Regularization$ & 0.1   \\ \hline
$BatchSize$ & 10  \\ \hline
$learning\_rate$ & 0.001   \\ \hline
$LearnRateDropFactor$ & 0.1   \\ \hline
$LearnRateDropPeriod$ & 20  \\ \hline
$LearnRateSchedule$ & $piecewise$   \\ \hline
$MaxEpochs$ & 80   \\ \hline
$Optimizer$ & SGD  \\ \hline
\end{tabular}}
\label{table3}
\end{table}

The model was trained using the stochastic gradient descent optimizer (SGD) with the help of the cross entropy loss given by:
\begin{equation}
    \label{crossentopyloss}
   L=\mathrm{-} \sum_{i=1}^{n} t_i\times log(p_i),
\end{equation}
where $t_i$ is the true label and $p_i$ is the softmax probability for the $ith$ class and $n$ is the number of classes.

\subsection{Results of the Proposed Framework}
\begin{figure}[h!]
\centering
\includegraphics[width=7.25in,height=3.25in]{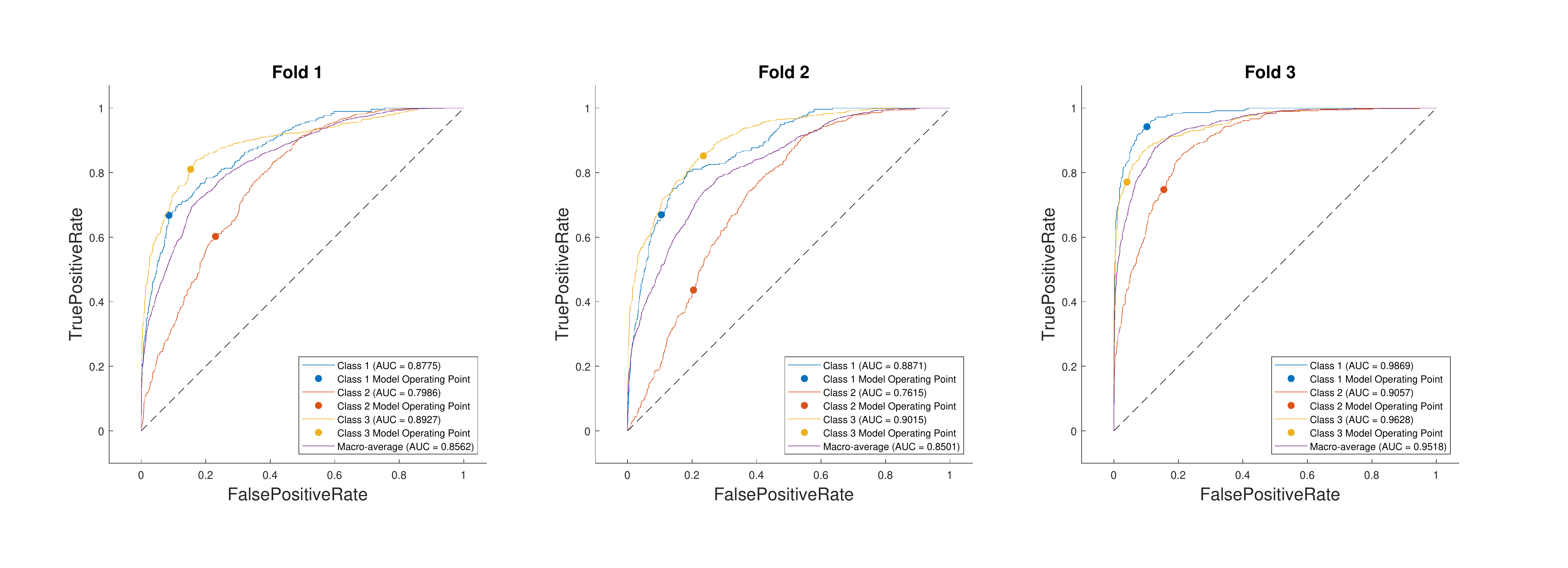}
\caption{Receiver-operating characteristic curves displaying the ability of the proposed unified model to perform the desired classification task in a 3-fold cross-validation setup  for PPG signal.}
\label{fig11}
\end{figure}

\begin{figure}[h!]
\centering
\includegraphics[width=7.25in,height=3.25in]{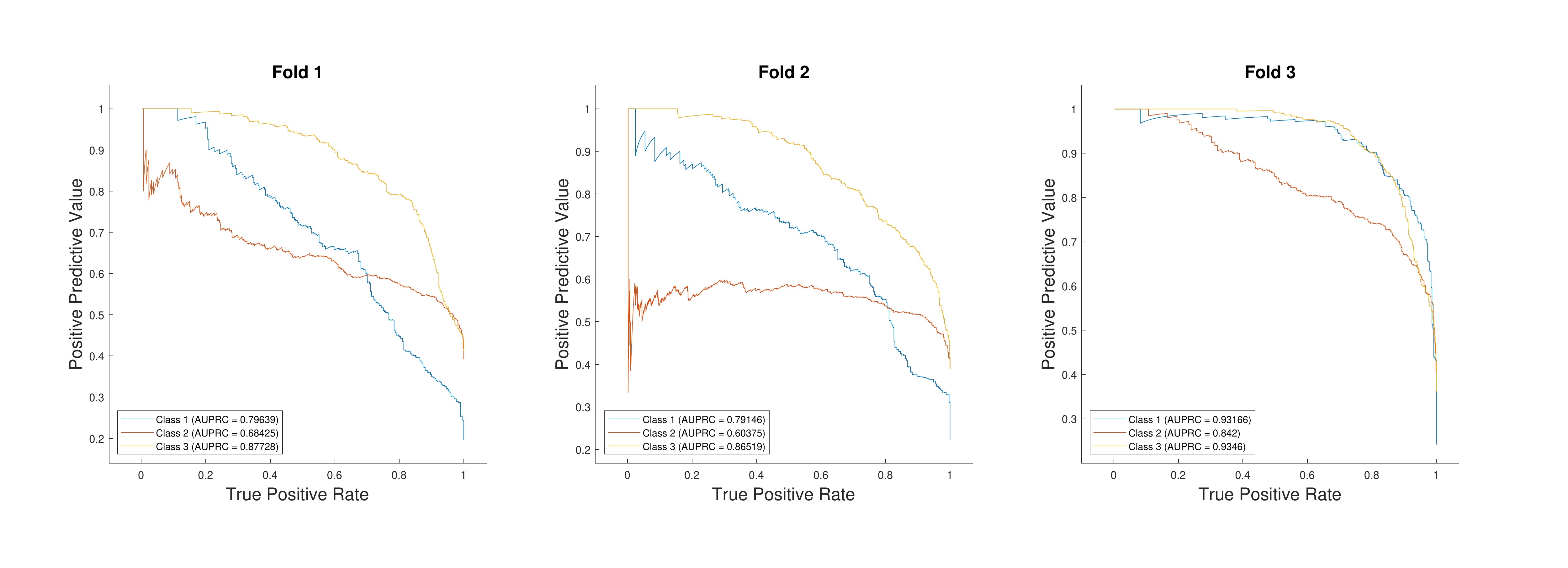}
\caption{Precision-Recall curves displaying the ability of the proposed unified model to perform the desired classification task in a 3-fold cross-validation setup for PPG signal.}
\label{fig12}
\end{figure}

Tables \ref{table4} and \ref{table5}  present the classification performance of the proposed method in a 3-fold cross-validation setup for both non-invasive signals, ECG and PPG respectively. The experimental results are initially evaluated and verified using the area under the receiver operating characteristic (AUROC) curve analysis in one vs others format for multi-class scenario. As an illustration, the individual AUROCs for each class with their corresponding model operating point, together with the average AUROC value for each fold, in a 3-fold cross-validation setup for PPG signal is shown in Figure \ref{fig11}. However, there exists a severe imbalance in data points among the three classes in terms of the available number of  waveform segments. Hence, average precision-recall curves are also analyzed as shown in Figure \ref{fig12}. Further, the harmonic mean between precision and recall, i.e.,  $F1$-score is also presented in Tables \ref{table4} and \ref{table5}.

\begin{table}[h!]
\small
\caption{Summary of cross-validation results for the proposed method and the ablation experiments on ECG signal.}

\label{table4}
\centering
\scalebox{.8}{
\begin{tabular}{|c|c|c|c|c|c|c|}
\hline
Models & \textbf{T-F moments} &  \textbf{Log-spectrograms}  &  \textbf{AUROC} & $F1_{score}$ & \textbf{Sensitivity} & \textbf{Specificity} \\ 
\hline
Branch 1 & \checkmark  &  \xmark  &0.6732 & 53.12  &61.14 &  69.05                        \\\hline
Branch 2 & \xmark  & \checkmark  &0.6432 & 49.12  &57.14 &  61.05                        \\\hline
\textbf{Proposed Study}  & \checkmark  & \checkmark  & \textbf{0.6953} & \textbf{ 56.67}  & \textbf{59.45} & \textbf{69.77 }  \\ 

\hline
\end{tabular}}
\end{table}

\begin{table}[h!]
\small
\caption{Summary of cross-validation results for the proposed method and the ablation experiments on PPG signal.}

\label{table5}
\centering
\scalebox{.8}{
\begin{tabular}{|c|c|c|c|c|c|c|}
\hline
Models & \textbf{T-F moments} &  \textbf{Log-spectrograms}  &  \textbf{AUROC} & $F1_{score}$ & \textbf{Sensitivity} & \textbf{Specificity} \\ 
\hline
Branch 1 & \checkmark  &  \xmark  &  85.64 & 68.81  &67.74 &  84.90 \\ 
\hline
Branch 2 & \xmark  & \checkmark  & 84.72 & 67.98  & 66.22&83.45 \\ 
\hline
\textbf{Proposed Study}  & \checkmark  & \checkmark  & \textbf{0.8861} & \textbf{72.16}  & \textbf{79.06} & \textbf{89.21}  \\ 
\hline
\end{tabular}}
\end{table}

\subsection{Ablation Experiments}
Further, to justify and emphasize the clinical performance of the proposed unified network architecture, a subjective analysis of the latent space derived  from the late-fusion at the end of DL unified model is done against the ablation experiments. These ablation studies involve testing the performance of the individual model in a respective  branch on the desired classification by excluding the  computation of  other branch model.

As a part of the aforementioned ablation experiments, we performed two well-tuned individual model training methods. In the first method, the model  trained only with T-F moments from branch 1 of Figure \ref{fig6}  is considered excluding the effect of \textit{'log spectrum layer'} from branch 2 of the unified DL model. For the latter ablation method the experimental setup is vice versa. The results for these ablation experiments are also presented in Tables \ref{table4} and \ref{table5} for ECG and PPG signals respectively. Model hyper-parameters  were always consistent across these evaluations as enlisted in Table 3. As seen in Tables \ref{table4} and \ref{table5}, the performance of the proposed unified model alone with  late fusion  was significantly higher in comparison to the individual ablations and further, the PPG signal outperformed ECG using the proposed unified model.

\subsection{Comparative Experimental Analysis}
The most commonly studied theories in the research context of  artificial distinction within simulated hemorrhage are arterial waveform analysis, which explores features in terms of fiducial points obtained from PPG derivatives and the heart (pulse) rate variability i.e., HRV (PRV)-based features from ECG and (or) PPG. Further, PPG morphological theory is also studied that mainly explores PPG signals. As mentioned earlier in  
$section$ \textit{2.3.3} this study used raw waveform segments and explored time-frequency representations to capture the transient signatures for the desired classification task. The reason behind resorting to T-F representations was to eliminate the limitations (mentioned in the introduction) exhibited by the cumbersome feature extraction involved in classical methods of AWFA (PPG morphological theory, HRV, PRV and artery wave propagation theory).

To support our hypothesis that the proposed T-F-based DL approach on given waveform segments is more efficient than the above-mentioned classical methods, we perform the comparative analysis with the following baseline studies w.r.t. two noninvasive modalities i.e. ECG and PPG. 

\subsubsection{ECG analysis}
\begin{itemize}
	    \item HRV features: Inter-beat-interval information is obtained by calculating the RR interval of each waveform segment \cite{ahmed2010heart}. Pan-Tompkins algorithm \cite{pan1985real,warlar1991integer} is employed to efficiently detect QRS wave and subsequently the R peaks. HRV-based features in time and  frequency-domain  are extracted from the derived RR intervals \cite{shaffer2017overview}. More details on HRV based features are presented in Table A1 of Appendix A.
	     \item Nonlinear features: Several nonlinear signatures from ECG were extracted. Entropy-based features inspired from the research Entropy-hub \cite{flood2021entropyhub}, auto-regressive coefficients from the model with order 4 \cite{zhao2005ecg}, Shannon entropy values from the sub-bands resulting using level 4 decomposition on maximal overlap discrete wavelet packet transform \cite{li2016ecg}, and Multifractal wavelet-based features of the scaling exponents were also extracted \cite{leonarduzzi2010wavelet}. These nonlinear features were chosen as per the  state-of-the-art research demonstrating their effectiveness in various ECG classification tasks. Detailed explanation of the extracted nonlinear features is presented in appendix A.
	     


	     \item Time-frequency analysis: Various T-F representations viz., scalograms, spectrograms, and wavelet scattering etc., were also explored to capture the transients of the ECG signal waveform segments. However, the scalogram-based CNN-LSTM DL model yielded better performance among T-F analysis. 
	\end{itemize}

\subsubsection{PPG analysis}
\begin{itemize}
	    \item Fiducial points-based features: A comprehensive investigation in terms of morphological characteristics of the PPG signal and its derivatives was carried out according to the recommendation in the research \cite{chen2020estimating,elgendi2018toward}. The detection of the fiducial points assists in extracting time, amplitude, locations, and finally morphological area of the  underlying signal. These signatures have been widely employed for tracking hemodynamics \cite{miao2017novel,miao2019multi}. These  investigated features  are detailed in Table A2 of Appendix A.
	     \item Pulse-rate variability (PRV) and  Nonlinear features: Similar to ECG-based  HRV analysis, PPG-based PRV features, and other nonlinear features were extracted.
	
	\end{itemize}	

\begin{table}[ht]
\tiny
\renewcommand{\arraystretch}{1.0}
\caption{Summary of comparative analysis for the classification performance between the proposed  method and  baseline studies in 3-fold cross-validation setup.}
\label{table6}
\centering
\begin{tabular}{|c|c|c|c|c|c|c|}
\hline
\begin{tabular}[c]{@{}c@{}c@{}} \textbf{Method}  \\ \textbf{used} \end{tabular} & \textbf{(\#)Features} & \textbf{Classifier} & \textbf{AUROC} & $F1_{score}$ & \textbf{Sensitivity} & \textbf{Specificity} \\ \hline

\begin{tabular}[c]{@{}c@{}c@{}} ECG analysis  \\ (HRV + nonlinear)  \end{tabular}  &  55 & Bag Decision Trees & 0.6432 & 49.12  &57.14 &  62.05                        \\\hline
\begin{tabular}[c]{@{}c@{}c@{}} ECG analysis  \\Time-frequency  \end{tabular}  & 2D Scalograms & CNN-LSTM & 0.7778& 61.64  & 65.07&81.09                          \\\hline
\begin{tabular}[c]{@{}c@{}c@{}}  PPG analysis  \\(Fiducial-points \\ Morphological)  \end{tabular}  & 48 & AdaBoost &0.7664 & 58.62  &57.74 &      79.13                    \\\hline
\begin{tabular}[c]{@{}c@{}c@{}}  PPG analysis  \\ (PRV + nonlinear)  \end{tabular}  & 55 &Bag Decision Trees &0.6928 & 55.67  &59.92 & 69.72                         \\\hline
\begin{tabular}[c]{@{}c@{}c@{}}  \textbf{Proposed Study}  \\ \textbf{(PPG)}  \end{tabular}  & \begin{tabular}[c]{@{}c@{}c@{}}  \textbf{T-F moments +}  \\ \textbf{log-spectrograms}  \end{tabular} & \textbf{Unified DL Model} & \textbf{0.8861} & \textbf{72.16}  & \textbf{79.06} &  \textbf{89.21}                        \\\hline

\end{tabular}
\end{table}

Table \ref{table6} represents the comparative analysis of the classification performance between the proposed method and the baseline studies. Table \ref{table6} shows that the classification performance of PPG-based T-F representations with the proposed unified model outperformed ECG signal and also significantly higher than other baseline studies performed  with classical methods for ECG and PPG. It is worth mentioning that the proposed unified framework with the same hyper-parameters enlisted in Table \ref{table3} with appropriate modifications in network architecture was applied  on the underlying classical methods for ECG and PPG.  However, the proposed model did not perform better compared to the  respective classifiers listed in Table \ref{table6}. This means only the results of the best performing classifiers are shown in Table \ref{table6}. 



\section{Discussion}\label{sec5}
The ternary classification results obtained from the proposed study demonstrate the possibility for the design of ML models using non-invasive waveforms to classify the level of hypovolemia prior to overt hemodynamic decompensation in healthy volunteers undergoing LBNP.  The unique transient signatures captured and learned by the proposed unified model from the raw waveform data are quite efficient compared to the classical morphological features.  This is hypothesized by our comparative results  of the classification performance between the proposed versus the classical feature extraction techniques.

ML techniques in comparison to statistical methods are data-driven and impart a comprehensive way toward reliable  diagnosis and prognostication. Contemporary research studies on hemodynamic monitoring  and its management strategies for the diagnosis of blood loss have been widely addressed by the development of such ML models. 
However, in retrospect, providing a  straightforward and direct comparison among these ML-based studies on hemodynamic instability is a tedious task because of certain reasons. viz., In these studies, generally the context of each problem to be addressed is varied. Some of the studies  demonstrated that,  forecasting the trend of certain vital signs by the ML models, learned with the initial partial part, can herald the condition of hemodynamic instability. Few of other studies tried to translate the model outputs into categories of physiological events, merely to have an artificial distinction for the classification task. Even variability exists among the experimental setup in terms of signal acquisition. Further, for the validation different performance metrics are used.

\begin{table}[!hbp]
\tiny
\renewcommand{\arraystretch}{1.0}
\caption{Summary of ML-based clinical studies performed using LBNP for  automated detection of simulated hemorrhage.}
\label{table7}
\centering
\begin{tabular}{|c|c|c|c|}
\hline
\textbf{\begin{tabular}[c]{@{}c@{}c@{}} Research  \\ Study\end{tabular}} & \textbf{\textbf{\begin{tabular}[c]{@{}c@{}c@{}} Methods  \\ (Features)\end{tabular}}} & \textbf{Model} & \textbf{Results} \\ \hline
Convertino \textit{et al.} \cite{convertino2011use} &\begin{tabular}[c]{@{}c@{}c@{}}Non-invasive hemodymaic features:\\BP,  $EtCO_2$, pulse
character, and \\respiratory rate \end{tabular}&\begin{tabular}[c]{@{}c@{}c@{}} Regression Analysis  \\ Leave-one-subject-out strategy\end{tabular} &\begin{tabular}[c]{@{}c@{}c@{}}Accuracy: 96.50 \% \\ Correlation Coefficient($R$): 0.89
 \end{tabular}\\\hline
Convertino \textit{et al.} \cite{convertino2022ai,techentin20191d} & CRI \& CRM & \begin{tabular}[c]{@{}c@{}c@{}}Logistic Regression Analysis, CNN\\ 10 \% hold out validation  \end{tabular} &\begin{tabular}[c]{@{}c@{}c@{}}Binary classification \\ AUROC: 0.9268 (CRM)\\ AUROC: 0.9164 (CRI)  \end{tabular} \\\hline
 Bjorn J.P.van der \textit{et al.} \cite{van2018support}        & \begin{tabular}[c]{@{}c@{}c@{}c@{}c@{}}BP Curve dynamics, SV, CO, \\
 $EtCO_2$ , TCD\\ \\  \end{tabular}       &\begin{tabular}[c]{@{}c@{}c@{}} SVM  \\ Leave-one-subject-out strategy\end{tabular}          & \begin{tabular}[c]{@{}c@{}c@{}}3-class classification study
\\Sensitivity: 78.21 \%
\\Specificity: 91.51 \% \end{tabular} \\ \hline
Bjorn J.P.van der \textit{et al.} \cite{https://doi.org/10.14814/phy2.13895} & \begin{tabular}[c]{@{}c@{}c@{}}1-D Cubic Hermite splines \\ interpolation + PCA \end{tabular} &\begin{tabular}[c]{@{}c@{}c@{}} SVM  \\ Leave-one-subject-out strategy\end{tabular} & \begin{tabular}[c]{@{}c@{}c@{}}
4-class classification study \\Accuracy: 57 \% \\ MSE: 0.26 \\Kappa: 0.4650
\end{tabular}\\\hline
Soo-Yeon Ji \textit{et al.} \cite{ji2013heart}& \begin{tabular}[c]{@{}c@{}c@{}} HRV analysis
\\ + Wavelet Transformation\end{tabular}   & \begin{tabular}[c]{@{}c@{}c@{}} LibSVM  \\ Leave-one-subject-out strategy\end{tabular}  & \begin{tabular}[c]{@{}c@{}c@{}}
Binary classification \\Accuracy: 89.1 \% \\ AUROC: 0.86 \\ 3-class classification study \\Accuracy: 69.5 \% 
\end{tabular} \\\hline
\textbf{Proposed Study}         &  \begin{tabular}[c]{@{}c@{}c@{}}Time-Frequency Moments
\\ + Log Spectrograms
 \end{tabular}   &\begin{tabular}[c]{@{}c@{}c@{}}Unified DL Model\\
3-fold cross-validation
 \end{tabular} & \begin{tabular}[c]{@{}c@{}c@{}}AUROC: 0.8861
 \\AUPRC: 0.8141
 \\$F1_{score}$: 72.16
\\Sensitivity: 79.06 \%
\\Specificity: 89.21 \% \end{tabular}     \\ \hline           
\end{tabular}
\end{table}

Table \ref{table7} summarizes ML-based clinical studies employing LBNP for automated detection and classification of simulated hemorrhage. A research study \cite{convertino2011use} led by Convertino \textit{et al.} developed a novel ML-based experimental setup using LBNP to estimate CBV loss  with 96.5\% accuracy. The correlation between actually applied LBNP levels and the  prediction for hemodynamic decompensation using forecasting was 0.89. Non-invasive  hemodynamic features were used in the design of ML model that mainly includes vital signs during surgery viz., blood pressure, $EtCO_2$, pulse character, and respiratory rate. More specific to the topic of classification among physiological events under simulated hemorrhage via LBNP experimental settings, the same group by Convertino \textit{et al.} performed a binary classification between Low versus High tolerance categories towards reductions in CBV \cite{convertino2022ai,techentin20191d}. A Logistic regression analysis by regressing the onset of decompensated shock was performed using two unique compensatory reserve algorithms viz., CRM (compensatory reserve metric) and CRI (compensatory reserve index) that yielded performance of AUROCs to be 0.9268 and 0.9164 respectively. Bjorn J.P.van der \textit{et al.} \cite{van2018support} developed a Support vector machine (SVM) - based predictive algorithm to perform ternary classification of impeding simulated hypovolemic  shock using LBNP. The model features included , BP curve dynamics, volumetric hemodynamic parameters (both SV and CO), $EtCO_2$, and middle cerebral artery transcranial Doppler (TCD) blood flow velocity. The average sensitivity and specificity for the ternary classification using \textit{'leave-one-subject-out'} validation were 78.21\% and 91.51\% respectively. Further, Bjorn J.P.van der \textit{et al.}  also reported results for a 4-class classification study \cite{https://doi.org/10.14814/phy2.13895} with accuracy, mean square error and Kappa score of 57\%, 0.26, and 0.4650 respectively. HRV analysis using ECG was done by Soo-Yeon Ji \textit{et al.}  \cite{ji2013heart} by applying wavelet-based ML predictive algorithms for the prediction of induced central hypovolemia via LBNP as a surrogate of hemorrhage. The average accuracy and AUROC for binary classification were 89.1 \% and 0.86  respectively and for ternary classification  accuracy of  69.5 \% is obtained using leave-one-subject-out validation.

All of the clinical studies existing to date, for the simulated hemorrhage deployed LBNP experimental setups based on  negative pressure that progressively descends step-wise with equal duration at each level and thus make the event to be biased with time dependency, with the hypothesis that as time elapses, there is continuous bleeding.

The proposed dynamic LBNP protocol with added randomness in LBNP levels and duration of levels was to reduce the effect of time and to mimic a more relevant clinical scenario where a bleeding patient receives fluid resuscitation from health personnel at the site of the accident or in the ambulance on the way to the hospital. Fluid resuscitation changes central blood volume. Typically, it is done intermittently with various rates, volumes and times, depending on a subjective evaluation of both the amount of the bleeding and the effects of interventions to stop the bleeding. As a consequence, the fluctuations in central blood volume can be large and rapid as simulated in our model. The results of our study were based on ML analysis of routine non-invasive vital signs as used in an ambulance. ECG and PPG are the only continuous monitoring modalities in this setting. An important finding was that the PPG signal performed better than ECG in classifying levels of bleeding. This means that reliable monitoring of changes in central blood volume is possible by solely using a finger probe. PPG waveform contains information on heart rate, and pulsatile volume in addition to arterial oxygenation of the patient. Respiration, sympathetic nervous system activity, and thermoregulation also influence the waveform. Our proposed ML algorithm captured this complex interplay in physiological responses to different levels of bleeding. From a clinical point of view, in the case of bleeding and fluid resuscitation, a method should also provide reliable information on treatment effects in order to avoid serious complications related to fluid overload. Our proposed ML algorithm on PPG signal demonstrates this possibility in this challenging LBNP setup. The PPG signal is usually of high quality in comparison to ECG, even during movement of the patient. This makes it likely that the ML algorithm can provide accurate early recognition and analysis of bleeding.


 Some of the limitations of our proposed study that need to be explored further in prospective real-time clinical deployment are as follows.
 Healthy subjects who were exposed to negative pressure via the experimental setup, mimic similar physiology of hemodynamics to that of subjects undergoing hemorrhage. However, they differ from the ones with actual bleeding because they were neither in pain, nor anxiety nor they had a disruption in actual tissues. Hence the complete translation of this experimental model to the actual scenario of patients in trauma with hemorrhage shock is not true. However, the proposed experimental setup imparts a peerless monitoring  opportunity of vital physiological changes in real-time that can map the compensatory responses to progressive central hypovolemia similar to that caused by bleeding. 
 Even, previous LBNP trials  exhibited identical physiologic responses to those of actual volume loss during the early compensatory phases of hemorrhage \cite{cooke2004lower,convertino2001lower,summers2009validation}.

Distribution of the waveform segments among the three classes highly varied as input observations to the model. The \textit{'mild'} (Class 1) blood loss  had a fewer observations. compared to \textit{'moderate'} and \textit{'severe'} blood loss classes. In case if the model was trained on relatively equal number of waveform segments among the three classes, then there could have been significant effect on the classification performance. However, the models trained with imbalanced data are still obliged to assign prediction labels to each waveform segment through the whole LBNP trajectory from normal- to hypovolemia among the three classes. This reflects in the output with the overall high specificity, since the miss-classification of an observation w.r.t., investigated class could result in observation belonging to  either of the two remaining classes.

There is no straightforward linkage in mapping the physiological events comprising of given LBNP trajectory, translating to the output of the DL-based AI model. i.e., three class definitions from normal- to hypovolemia, were created  merely to accomplish an artificial distinction among the ongoing hemodynamic decompensation towards progressive central hypovolemia. From a clinical perspective, it is still debatable to comment that the underlying physiological responsive events may fit (or) not fit into these classes, and hence classification performances thus reported may not reflect the direct classification of underlying physiology. However, changing probabilities by the DL model among the class definitions quantify model performance  that hints at the progression of hemodynamic instability respectively.

\section{Conclusion and Future work}
In this study, a novel DL-based model using a modified dynamic LBNP expiremental setup is developed to explore the strength of T-F representations on non-invasive waveforms for the classification of hemodynamic decomposition levels. The experimental findings showed that the PPG waveform induced detectable changes compared to ECG  for blood loss among awake and spontaneously breathing subjects. These changes that were earlier elusive to clinicians might be now captured by T-F analysis, in which we show that spectral amplitudes via center moments and log-scale spectrum correlate to blood volume loss. Thus a computer-aided diagnostic algorithm coupled with novel DL techniques  can be designed to monitor non-invasive pulse waveforms which is capable of identifying the correlation of ongoing blood volume loss within them. This will certainly prove critical in treating hemorrhage and avoiding likely episodes of irreversible shock. This DL framework is an initial promising approach in the direction of improving the clinical outcomes for combating casualties of patients with unrecognized hemorrhage and other forms of impending hemodynamic instability.

Future efforts will be to emphasize the ML models’ discriminative ability in a varied collection of data under different human experimental protocol settings. We understand that the cohorts under study were limited only to healthy volunteers and hence experimental protocols with various patient conditions will allow the model to \textit{“learn”} to be more diagnostic. We also plan to explore the strength of time-series transformers on the underlying non-invasive waveforms to eliminate the need for feature extraction and strengthen the model with a multi-head attention mechanism.

\section*{Conflicts of Interest}
The authors declare no conflict of interest.
\section*{Funding}  
This work was supported in part by the Health South East Authority in Norway, Helse Sør-Øst RHF (HSØ: \textit{New Real-time Decision Support during Blood Loss using Machine Learning on Vital Signs}) under Grant no. 19/00264-202, and Prosjektnummer 2020079.

\section*{Ethical Approval} 
The study was approved by the regional ethics committee (REK sør-øst C/ 2019/ 649). After written informed consent, 23 healthy volunteers aged between 18 and 40 years were included in the study. Pregnancy and/or cardiovascular disease with medication were exclusion criteria.
\section*{Acknowledgment}
The authors acknowledge the support provided for the experimental setup by Jonny Hisdal, from Faculty of Medicine, University of Oslo, and Section of Vascular Investigations, Oslo University Hospital, Oslo, Norway. Consultant Marius Erichsen from Division of Emergencies and Critical Care, Department of Anesthesiology, Oslo University Hospital, for including volunteers and participating in the LBNP experiments. Consultant Sverre Nestaas from Division of Emergencies and Critical Care, Department of Anesthesiology, Oslo University Hospital, for participating in preparing the dataset for analysis.

\bibliography{main}

\end{document}